\documentclass[prb,showpacs,superscriptaddress,twocolumn,amssymb,amsfonts]{revtex4-1}
\usepackage{amsmath}
\usepackage{bm}
\usepackage{epsfig}
\usepackage{graphics}
\usepackage{dsfont}

\def\beq{\begin{equation}}
\def\eeq{\end{equation}}
\def\beqn{\begin{eqnarray}}
\def\eeqn{\end{eqnarray}}

\begin{document}
\title{Superconductivity of doped Weyl semimetals: finite-momentum pairing and electronic analogues of the $^{3}$He-A phase}
\author{Gil Young Cho}
\affiliation{Department of Physics, University of California,
Berkeley, CA 94720, USA}
\author{Jens H. Bardarson}
\affiliation{Department of Physics, University of California,
Berkeley, CA 94720, USA}
\affiliation{Materials Sciences Division,
Lawrence Berkeley National Laboratory, Berkeley, CA 94720, USA}
\author{Yuan-Ming Lu}
\affiliation{Department of Physics, University of California,
Berkeley, CA 94720, USA} 
\affiliation{Materials Sciences Division,
Lawrence Berkeley National Laboratory, Berkeley, CA 94720, USA}
\author{Joel E. Moore}
\affiliation{Department of Physics, University of California,
Berkeley, CA 94720, USA} \affiliation{Materials Sciences Division,
Lawrence Berkeley National Laboratory, Berkeley, CA 94720, USA}


%

\begin{abstract}
We study superconducting states of doped inversion-symmetric Weyl semimetals. Specifically, we consider a lattice model realizing a Weyl semimetal with an inversion symmetry and study the superconducting instability in the presence of a short-ranged attractive interaction. With a phonon-mediated attractive interaction, we find two competing states: a fully gapped finite-momentum (FFLO) pairing state and a nodal even-parity pairing state.  We show that, in a BCS-type approximation, the finite-momentum pairing state is energetically favored over the usual even-parity paired state and is robust against weak disorder. Though energetically unfavorable, the even-parity pairing state provides an electronic analogue of the $^3$He-A phase in that the nodes of the even-parity state carry non-trivial winding numbers and therefore support a surface flat band. We briefly discuss other possible superconducting states that may be realized in Weyl semimetals.
\end{abstract}


\maketitle

\section{introduction}
Recent progress in our understanding of non-interacting Bloch electrons~\cite{classification1, classification2,  mb, fkm, hm, qi1} reveals a large class of gapped topological phases, the so-called topological insulators and superconductors~\cite{hm, QiRev, KaneRev}. For example, a time-reversal-symmetric topological insulator is a band insulator that cannot be continuously tuned into a trivial atomic insulator, as long as time-reversal symmetry is respected. A topological insulator is featured by a single Dirac cone in its surface state spectrum. Typically, these topological phases are realized in systems with strong spin-orbit coupling. It is known that when topological insulators are combined with superconductivity via the proximity effect~\cite{fu1} or via phonon-mediated attractive interaction~\cite{fu2}, the interesting interplay between electron pairing and the spin-orbit coupling results in exotic superconductivity. For instance, when a topological insulator is doped and turned into a superconductor, an odd-parity topological superconductor is obtained~\cite{fu2} with a single gapless Majorana surface state, which is protected by time-reversal symmetry.

Another type of gapless `topological matter', the Weyl semimetal, is currently being studied intensively and is proposed to be realized in experiments~\cite{Ashvin, Burkov, Yang, cho, latticeWeyl}. Its electronic structure has an {\it even} number of Weyl nodes -- two cylyndrical 3D cones that touch at their apex, the Weyl point --  which carry non-trivial winding numbers ensuring their stability. These Weyl nodes can be thought of as 3D analogs of the two component Dirac fermions in graphene and at the surface of a 3D topological insulators.  They exhibit spin-momentum locking and thus require strong spin-orbit coupling to be realized. In analogy with superconductivity in topological insulators, it is natural to expect interesting superconducting states to emerge in these systems upon doping, resulting from their non-trivial topological winding numbers. To realize a Weyl semimetal phase requires either time-reversal symmetry~\cite{Ashvin} or inversion symmetry~\cite{murakamiweyl} to be broken.  In this paper we concentrate on the inversion-symmetric case, in which the two nodes connected by the inversion symmetry carry opposite chirality. Upon slight doping there are at least two disconnected components to the Fermi surface around the nodes, shifted in momentum space from the inversion-symmetric high-symmetry points such as the $\Gamma$-point. We will show that the interplay between the finite-momentum displacement and the non-trivial winding numbers around each Weyl node leads to interesting superconducting states. 

The finite momentum shift of the Fermi surface motivates the study of finite-momentum pairing states or Fulde-Ferrell-Larkin-Ovchinnikov (FFLO) states~\cite{FF,LO}. FFLO states break translational symmetry and have interesting physical properties~\cite{FF,LO, disorder3}. In the Weyl semimetals, the center of momentum of the FFLO pairs is fixed by the momentum of the Weyl nodes. Similarly, the non-trivial winding around the nodes and the broken time-reversal symmetry suggests the possibility of realizing even/odd-parity BCS states that are electronic analogues of the $^{3}$He-A phase~\cite{He3a1,He3a2}. Since the $^{3}$He-A phase has nodes with non-trivial winding number which guarantees the existence of a dispersionless surface states~\cite{He3a2}, the Weyl semimetal in these phases is also expected to support zero-energy surface flat bands, similar to a Weyl semimetal in proximity to a superconductor~\cite{He3a2,WeylSc,Fa,nodal}.

Surprisingly, when the attractive interaction is completely local in real space and represents a phonon-mediated interaction, we find from a self-consistent mean-field calculation that the {\it fully-gapped} finite-momentum pairing is energetically favored over the even-parity BCS state (both pairing states can be thought of as spin-`singlet' pairings though `singlet' is not a very exact terminology since spin-rotational symmetry is broken) and is stable against weak disorder. Hence, there is a good chance of experimentally observing these exotic phases. To be concrete, we concentrate on a specific lattice model realizing an inversion-symmetric Weyl semimetal and solve the gap equation of the lattice model in the BCS approximation. We also discuss the applicability of our result to other models realizing Weyl semimetals.

The proximity effect of an s-wave superconductor on an undoped Weyl semimetal has been studied by Meng and Balents in Ref.~\onlinecite{WeylSc}. In contrast to this work, where the superconductivity is {\it extrinsic}, we are interested in the {\it intrinsic} superconductivity of the doped Weyl semimetal. Note that if the Weyl semimetal is undoped, the intrinsic superconducting gap and the critical temperature are expected to be vanishingly small since the density of states goes to zero at the Weyl point.

\section{Model}
The model we consider in this work is given by the Hamiltonian 
\begin{equation}
	H = H_0 + V_{\rm ee}
	\label{eq:Model}
\end{equation}
where $V_{\rm ee}$ is an electron-electron interaction term to be specified below. For the kinetic term, $H_0$, we take the minimal two-band lattice model~\cite{Yang} 
\begin{align}
H_0 = t& (\sigma^{x} \sin k_{x} + \sigma^{y} \sin k_{y})+t_{z}(\cos k_{z} - \cos Q)\sigma^{z}  \nonumber \\  &+m(2-\cos k_{x}-\cos k_{y})\sigma^{z} -\mu.
\label{lattice}
\end{align}
This model realizes a Weyl semimetal with two Weyl points at momenta ${\vec P}_\pm = (0,0,\pm Q)$. $\sigma^{x,y,z}$ are the Pauli sigma matrices (for later use we define $\sigma^0$ to be the $2\times2$ unit matrix), $t$ and $t_z \sin Q$ are the Fermi velocities at the Weyl points in the $x,y$ and $z$ directions respectively. Without a loss of generality, we assume $t = t_{z}\sin(Q)$ such that the Fermi velocity around the Weyl points is isotropic. We have explicitly included the chemical potential $\mu$ in the kinetic term. We are primarily interested in the parameter range $0 < |\mu/t| \ll Q$, when the Fermi surface consists of two disconnected spherical components around the Weyl points (see Fig.~\ref{Fig1}). In this case, the states on the Fermi surface have spin-momentum locking similar to the surface states of a strong topological insulator. This property will play an important role in our discussion of pairing states below.

The electron-electron interaction is short ranged and takes the form
\beq
V_{\rm ee} = V_{0} \sum_{i} n_{i}n_{i} + V_{1} \sum_{\langle ij\rangle}n_{i}n_{j} =  \sum_{{\vec k}} V({\vec k})n_{{\vec k}}n_{-{\vec k}},
\label{interaction}
\eeq
where $n_{i} = \sum_{\sigma} c^{\dagger}_{i, \sigma}c_{i,\sigma}$ is the number of electrons on site $i$, and the second sum is over nearest neighbors only. $V({\vec k}) = V_{0}+V_{1}(\cos k_{x} + \cos k_{y} + \cos k_{z})$ is the Fourier transform of the real-space interaction. $V_{0}$ represents a phonon-mediated attractive on-site interaction and $V_{1}$ the nearest-neighbor interaction. We are mainly interested in the case $V_{0}<0$ and $|V_0|\gg |V_{1}|$ when electrons form Cooper pairs and condense. The phenomenological interaction term~\eqref{interaction} captures the tendency of d-wave pairing for $V_{1}<0$ and $V_{0}>0$ in the context of high-Tc superconductors such as cuprates~\cite{MacDonald}. 

The point group symmetry of the model Hamiltonian, $H$, is $C_{4h} =\{I^{\eta_I}C_4^{\eta_4}|\eta_I=0,1;\eta_4=0,1,2,3\}$, where
\begin{align}
I&: \sigma^{z} H(-{\vec k})\sigma^{z} = H({\vec k}), \nonumber \\
C_{4}&: S^{\dagger} H[R_{\pi/2}({\vec k)}]S = H({\vec k}),\label{symmetry}
\end{align}
with $S = \frac{1}{\sqrt{2}}(\sigma^{0}+i\sigma^{z})$ and $R_{\pi/2}$ a rotation by an angle $\pi/2$ around the $z$-axis. $I$ is the inversion symmetry that takes ${\vec r} \rightarrow -{\vec r}$. Each spatial rotation is accompanied by an equal spin rotation, manifesting the spin-momentum locking due to the presence of strong spin-orbit coupling. The $C_4$-rotation symmetry in the $xy$-plane is not required to realize a Weyl semimetal, but is present in this model and similar lattice rotations are present in other models we discuss later. 

To make a connection to previous work~\cite{Ashvin, Yang, cho, Burkov} and to obtain a general understanding of Weyl semimetal phases, we derive the low energy effective theory corresponding to the lattice model~\eqref{eq:Model}. Expanding $H(\vec k)$ in the small momentum ${\vec q} = {\vec k} - {\vec P}_\pm$ around the two Weyl Points denoted by $\pm$, we obtain
\beq
H_0 = \sum_{{\vec k}}c^{\dagger}({\vec k}) H_0({\vec k}) c({\vec k}) \approx \sum_{a= \pm} \psi^{\dagger}_{a}({\vec q}) h_{a}({\vec q}) \psi_{a} ({\vec q}).
\label{weyl}
\eeq
The effective kinetic Hamiltonian $h_{\pm}({\vec q})$ is given by
\beq
h_{\pm}({\vec q}) = t (q_{x}\sigma^{x}+q_{y}\sigma^{y} \mp q_{z}\sigma^{z}) - \mu.
\label{low}
\eeq
Similarly, we obtain for the interaction term
\beq
V_{\rm ee} = \sum_{{\vec k},{\vec p}, {\vec q}}V^{ab;cd}({\vec q})\psi^{\dagger}_{a, \sigma}({\vec k}+{\vec q})\psi^{\dagger}_{b,\tau}({\vec p}-{\vec q})\psi_{c,\tau}({\vec p})\psi_{d,\sigma} ({\vec k}),
\eeq
where roman letters denote the nodal indices $\pm$ and $\sigma,\tau$ are spin indices. Here and henceforth, repeated indices are summed over. In the BCS channel (see App.~\ref{app:interactions} for details) 
\begin{equation}
V_{\rm ee} =\sum_{{\vec k},{\vec l}}V^{ab;cd}\psi^{\dagger}_{a, \sigma}({\vec k})\psi^{\dagger}_{b,\tau}(-{\vec k})\psi_{c,\tau}(-{\vec l})\psi_{d,\sigma} ({\vec l}),
	\label{eq:VeeBCS}
\end{equation}
with
\begin{align}
&V^{-+;+-}=V^{+-;-+}= V_{0} + 3V_{1} - \frac{V_{1}}{2}({\vec k}-{\vec l})^{2},\nonumber \\
&V^{-+;-+}= V_{0}+2V_{1} +V_{\perp} + V^{+}_{\parallel},\label{ContinuumInteraction}\\
&V^{+-;+-} =V_{0}+2V_{1} +V_{\perp} + V^{-}_{\parallel}, \nonumber 
\end{align}
and 
\begin{align}
V_{\perp} &= -\frac{V_{1}}{2} ({\vec k}_{\perp}-{\vec l}_{\perp})^{2},\\
V^{+}_{\parallel}/V_{1} &= [1 - \frac{1}{2}(k_{z}-l_{z})^{2}]\cos2Q + (k_{z}-l_{z})\sin2Q,\nonumber \\
V^{-}_{\parallel}/V_{1} &= [1 - \frac{1}{2}(k_{z}-l_{z})^{2}]\cos2Q - (k_{z}-l_{z})\sin2Q,\nonumber
\end{align}
where ${\vec k}_{\perp} = (k_{x},k_{y},0)$. These expressions will be used in the next section.

\section{Mean field theory and pairing channels}
\label{MFEnergy}
We treat the interaction term $V_{\rm ee}$ in a mean-field approximation and solve the resulting gap equations self-consistently. In addition to the more standard BCS paring, we also study finite-momentum or FFLO pairing.

\subsection{BCS pairing}
\begin{figure}
\includegraphics[width=1\columnwidth]{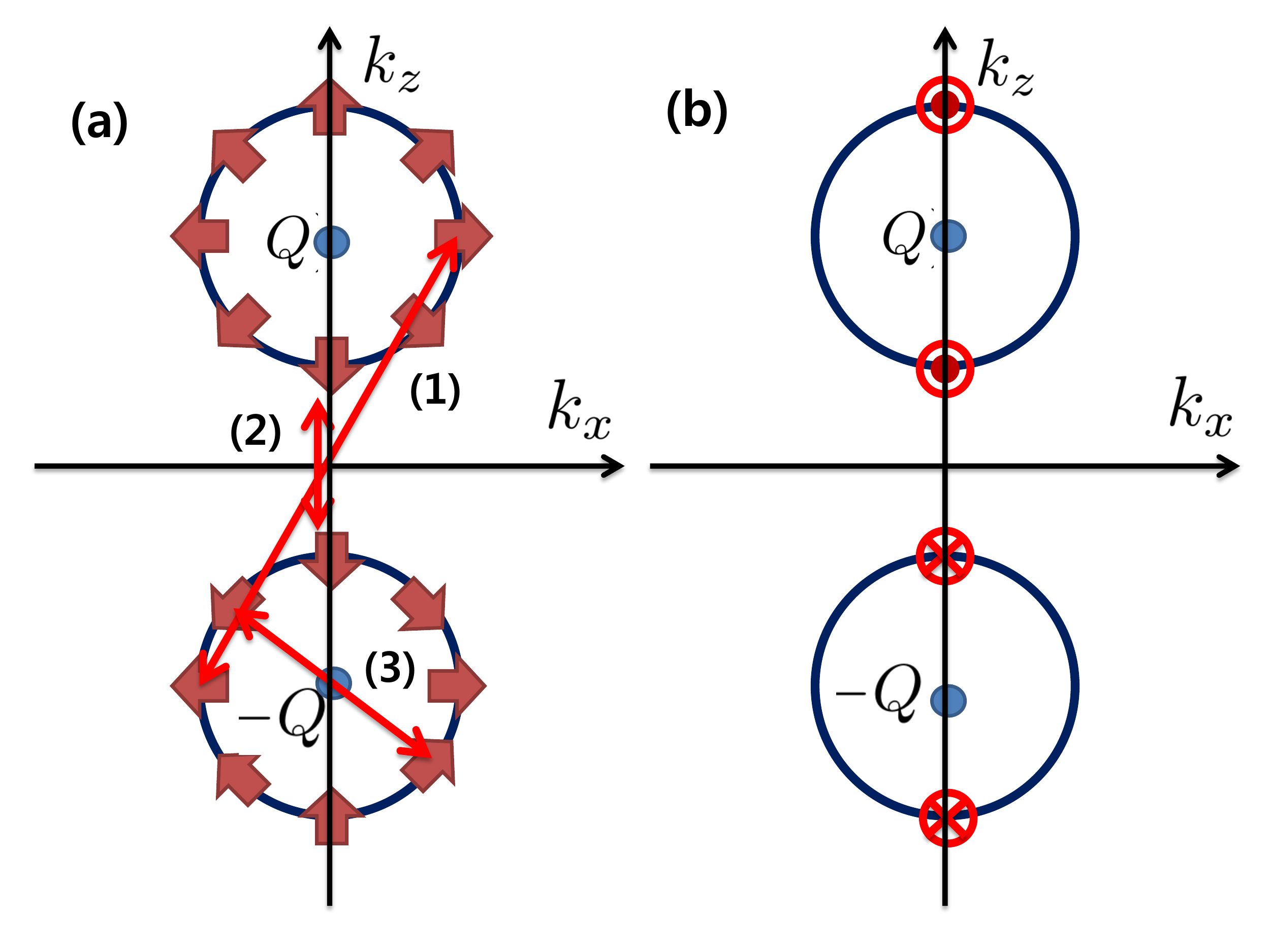}
\caption{A schematic diagram of the spin texture around the Weyl nodes in momentum space and the pairing states. (a) The spin direction of the eigenstates is given by thick arrows. The double-headed arrows labeled with (1) and (2) indicate the partner states in the BCS pairing. The spin state is maximally anti-parallel for (1) and parallel for (2), indicating that there will be nodes in the latter case if the pairing is in the singlet channel. Contrary to the BCS pairing, the FFLO pairing (3) connects two states within the same node (`intra-nodal' pairing). The two states connected by the FFLO pairing have the opposite spin directions. (b) Position of the nodes for the even-parity state. The nodes of the same chirality are on the same component of the Fermi surface, with their partner nodes of opposite chirality on the other. The filled circle represents a node of chirality $+1$ and the crossed circle represents a node of chirality $-1$. Hence, there are four nodal points on the Fermi surface of the even-parity paired BCS state.}
\label{Fig1}
\end{figure}

The symmetry classification of different BCS pairing order parameters in a doped Weyl semimetal, according to the lattice symmetry~\eqref{symmetry}, is summarized in Table \ref{table1} (see App.~\ref{app:symm} for more details). There are three fully-gapped BCS pairing order parameters ($\Gamma^1$ and $\Gamma^{3,\pm}$) and one that has nodal lines ($\Gamma^2$). 
\begin{table}
\begin{tabular} {l|l|l|l}
\hline
{}&IRR&$C_{4h}$&pairing function\\ \hline
$\Gamma^{1}$&$A$&$1$& $Z^2(i\sigma^{y}),(X-iY)(\sigma^0+\sigma^z),(X+iY)(\sigma^0-\sigma^z)$\\ \hline
$\Gamma^{2}$&$B$&$-1$&$XY(i\sigma^y),(X+iY)(\sigma^0+\sigma^z),(X-iY)(\sigma^0-\sigma^z)$\\ \hline
$\Gamma^{3,+}$&$E_{+}$&$i$&$(X+iY)Z(i\sigma^y),Z(\sigma^0+\sigma^z),XYZ(\sigma^0-\sigma^z)$\\ \hline
$\Gamma^{3,-}$&$E_{-}$&$-i$&$(X-iY)Z(i\sigma^y),XYZ(\sigma^0+\sigma^z),Z(\sigma^0-\sigma^z)$\\
\hline
\end{tabular}
\caption{Symmetry classification of the BCS pairing order parameters for the model (\ref{lattice}) according to the different irreducible representations (IRRs) of the group $C_{4h}$. $X$, $Y$, and $Z$ are basis functions for the momentum-space pairing function which we take to be $\sin p_{x}$, $\sin p_{y}$, and $\sin p_{z}$ respectively, and can be realized by nearest-neighbor pairings. Among the pairing functions, $i\sigma^y$ means singlet pairing, $\sigma^x$ is the spinfull triplet pairing, $\sigma^0+\sigma^z$ the triplet pairing for polarized $\uparrow$ spins and $\sigma^0-\sigma^z$ the triplet pairing for polarized $\downarrow$ spins. The paired state $\Gamma^2$ has nodal lines, while the other three states are gapless.}
\label{table1}
\end{table}

In the continuum theory, the pairing terms of Table~\ref{table1} take the form
\beq
\sum_{{\vec k}}\Delta_{\sigma\tau}({\vec k})c^{\dagger}_{\sigma}({\vec k})c^{\dagger}_{\tau}(-{\vec k})\approx \sum_{{\vec q}}\Delta^{a,b}_{\sigma\tau} ({\vec q}) \psi^{\dagger}_{a,\sigma}({\vec q})\psi^{\dagger}_{b,\tau} (-{\vec q}),
\eeq
The standard BCS pairing term connects two Weyl nodes in the effective theory. The explicit form of $\Delta^{a,b}_{\sigma\tau}$ and $\Delta_{\sigma\tau}$ can be found in Table \ref{table1} and in Eq.~\eqref{Pair} in App.~\ref{app:interactions}. The self-consistent gap equation takes the form
\beq
\Delta^{ab}_{\sigma\tau}({\vec p}) = \sum_{{\vec k}}V^{ab;cd}({\vec p} - {\vec k}) \langle\psi_{c,\tau}(-{\vec k})\psi_{d,\sigma}({\vec k})\rangle,
\eeq
where the expectation value is taken with respect to the mean-field superconducting state (see App.~\ref{app:interactions} for more explicit expressions for the gap equations).

\subsection{FFLO pairing}
In the doped Weyl semimetal, the Fermi surface is formed around the Weyl points ${\vec P}_\pm$, and it is natural to expect a finite-momentum pairing to compete with the standard BCS-paired states. We therefore introduce a FFLO state with a center of momentum at $2{\vec P}_\pm$, which paring function satisfies 
\beq
\Delta^{\pm}_{\rm FFLO}({\vec r}) \propto \exp(2i{\vec P_+}\cdot {\vec r}) \pm \exp(2i{\vec P}_-\cdot {\vec r})
\label{fflo}
\eeq
The self-consistent equations for these pairing order parameters take the form
\beq
\Delta_{\sigma\tau}({\vec p}; \pm {\vec P}) = \sum_{{\vec k}}V({\vec p} -{\vec k}) \langle \psi_{\pm,\tau}(-{\vec k})\psi_{\pm,\sigma}({\vec k})\rangle,
\eeq
where the two nodes $\pm$ are decoupled. These FFLO states correspond to the {\it intra-node} pairing, in contrast to the BCS case which is {\it inter-node} pairing (see Fig.~\ref{Fig1}). The two states of the pairings $\Delta^{\pm}_{\rm FFLO}$ in Eq.~\eqref{fflo} with a relative phase of $\pm1$ between the two components of the Fermi surface have the {\it same} mean-field energy since the two nodes are decoupled in the mean-field theory.
\begin{table}
\begin{tabular} {l|l|l}
\hline
{}&$C_{4}$&Pairing function\\ \hline
$\Gamma^{1}$&$1$& $Z^2(i\sigma^{y}),(X-iY)(\sigma^0+\sigma^z),(X+iY)(\sigma^0-\sigma^z)$\\ \hline
$\Gamma^{2}$&$-1$&$XY(i\sigma^y),(X+iY)(\sigma^0+\sigma^z),(X-iY)(\sigma^0-\sigma^z)$\\ \hline
$\Gamma^{3,+}$&$i$&$(X+iY)Z(i\sigma^y),Z(\sigma^0+\sigma^z),XYZ(\sigma^0-\sigma^z)$\\ \hline
$\Gamma^{3,-}$&$-i$&$(X-iY)Z(i\sigma^y),XYZ(\sigma^0+\sigma^z),Z(\sigma^0-\sigma^z)$\\
\hline
\end{tabular}
\caption{Classification of the FFLO states of superconducting Weyl fermions based on the lattice symmetry $C_{4}$. The notation is the same as in Table~\ref{table1}. We assume that the center of momentum for the pairing is at ${\vec P}_\pm$. Note that the symmetry is only $C_{4}$ on the $xy$-plane without the inversion because `inversion' is already encoded by the ansatz Eq.~\eqref{fflo}. This classification is essentially the same as that of BCS-type pairing order parameters.}
\label{table2}
\end{table}

\subsection{Mean field energy}
Having identified the possible superconducting states, we compute their free energy by solving the self-consistent gap equations numerically. We are interested in the case $|V_0|\gg |V_1|$ with $V_0 < 0$ where the spin singlet is preferred. We denote the pairing term $\propto i\sigma^y$ in Table~\ref{table1} as 'singlet' and the other terms $\propto i{\vec \sigma}\sigma^{y}$ as 'triplet'. The singlet and triplet components have a different dependence on the interaction parameters $V_{0}$ and $V_{1}$. The gap of the triplet components depends only on the value of $V_{1}$, while the singlet component depends only on $V_{0}+3V_{1} \approx V_{0}$ or $V_{0}+ V_{1}(5 + \cos 2Q)/2 \approx V_{0}$. We therefore consider in the following only the singlet component $\propto i\sigma^y$ of $\Gamma^1$ for the BCS and FFLO states. 

For these two states the BCS mean-field approximation is
\begin{equation}
	H = H_0 + V_{\rm ee}^{\rm pair},
	\label{eq:Hmf}
\end{equation}
where $H_0$ is given by Eq.~\eqref{lattice} and $V_{\rm ee}^{\rm pair}$ is the effective {\it projected} pair potential derived from the lattice interaction in Eq.~\eqref{interaction}. For the $\Gamma^{1}$-BCS state we have 
\begin{align}
V_{\rm ee}^{\rm pair} &= -U_{\rm BCS}\sum_{{\vec k}, {\vec p}}P^{\dagger}_{{\vec k}}P_{-{\vec p}}, \notag \\		
U_{\rm BCS} &= V_{0}+V_{1}\frac{5 + \cos(2Q)}{2},\\
P^{\dagger}_{{\vec k}} &= \psi^{\dagger}({\vec k}) \tau^{x}i\sigma^{y} \psi^{*}(-{\vec k}), \notag
\end{align}
and the gap equation is 
\beq
\Delta = -\frac{U_{\rm BCS}}{4} \int_{{\vec k}} \langle\psi_{a,\alpha}({\vec k}) (\tau^{x})^{ab}(-i\sigma^{y})^{\alpha\beta}\psi_{b,\beta}(-{\vec k})\rangle.
\eeq
For the $\Gamma^{1}$-FFLO state we obtain
\begin{align}
V_{\rm ee}^{\rm pair} &= -U_{\rm FFLO}\sum_{{\vec k}, {\vec p}}P^{\dagger}_{{\vec k}}P_{{\vec p}}  \notag \\
U_{\rm FFLO} &= V_{0}+3V_{1}, \\
P^{\dagger}_{{\vec k}} &= \psi^{\dagger}({\vec k}) i\sigma^{y} \psi^{*}(-{\vec k}), \notag 
\end{align}
with a gap equation 
\beq
\Delta =  -\int_{{\vec k}} \frac{U_{\rm FFLO}}{2} \langle\psi_{a,\alpha}({\vec k})(-i\sigma^{y})^{\alpha\beta}\psi_{a,\beta}(-{\vec k})\rangle
\eeq

In this standard BCS-type approximation, we can evaluate the energy $E$ of the pairing states with the pairing amplitude $\Delta_{\vec k} = -U\langle P_{-{\vec k}}\rangle$ (with the effective pairing interaction strength $U$) by computing
\begin{align}
E &= E_{\rm el} + E_{\rm sc}, \nonumber\\
E_{\rm el} &= \sum_{{\vec k} \in d\Omega, \epsilon_{\rm sc} <0} \epsilon_{\rm sc}({\vec k})n_{e}[\epsilon_{\rm sc}({\vec k})] \nonumber \\
& -\sum_{{\vec k} \in d\Omega, \epsilon_{\rm fs}<0} \epsilon_{\rm fs}({\vec k})n_{e}[\epsilon_{\rm fs}({\vec k})], \nonumber\\
E_{\rm sc} &= - \sum_{{\vec k} \in d\Omega}\frac{\Delta_{{\vec k}}\Delta^{*}_{-{\vec k}}}{2U} + h.c.
\label{Energy}
\end{align}
Here $n_{e}$ is the filling of the electron for the state at ${\vec k}$ of energy $\epsilon({\vec k})$, $\epsilon_{\rm sc}$ is the energy of the filled band of the BdG quasiparticle with mean-field gap $\Delta$, and $\epsilon_{\rm fs}$ is the energy of the filled bands of the free Weyl electrons without pairing, i.e.\ the energy of the normal state. Thus, the second line of Eq.~\eqref{Energy} represents the energy  gain of the superconducting state relative to the normal state by opening up a gap near the Fermi surface. The last line of Eq.~\eqref{Energy} represents the contribution from the pairing interaction labeled by the momentum ${\vec k}$. The range of the summation is restricted to a shell $d\Omega$ around the Fermi surface, which width is determined by the strength of the attractive interaction. 

\begin{figure}
\includegraphics[width=1\columnwidth]{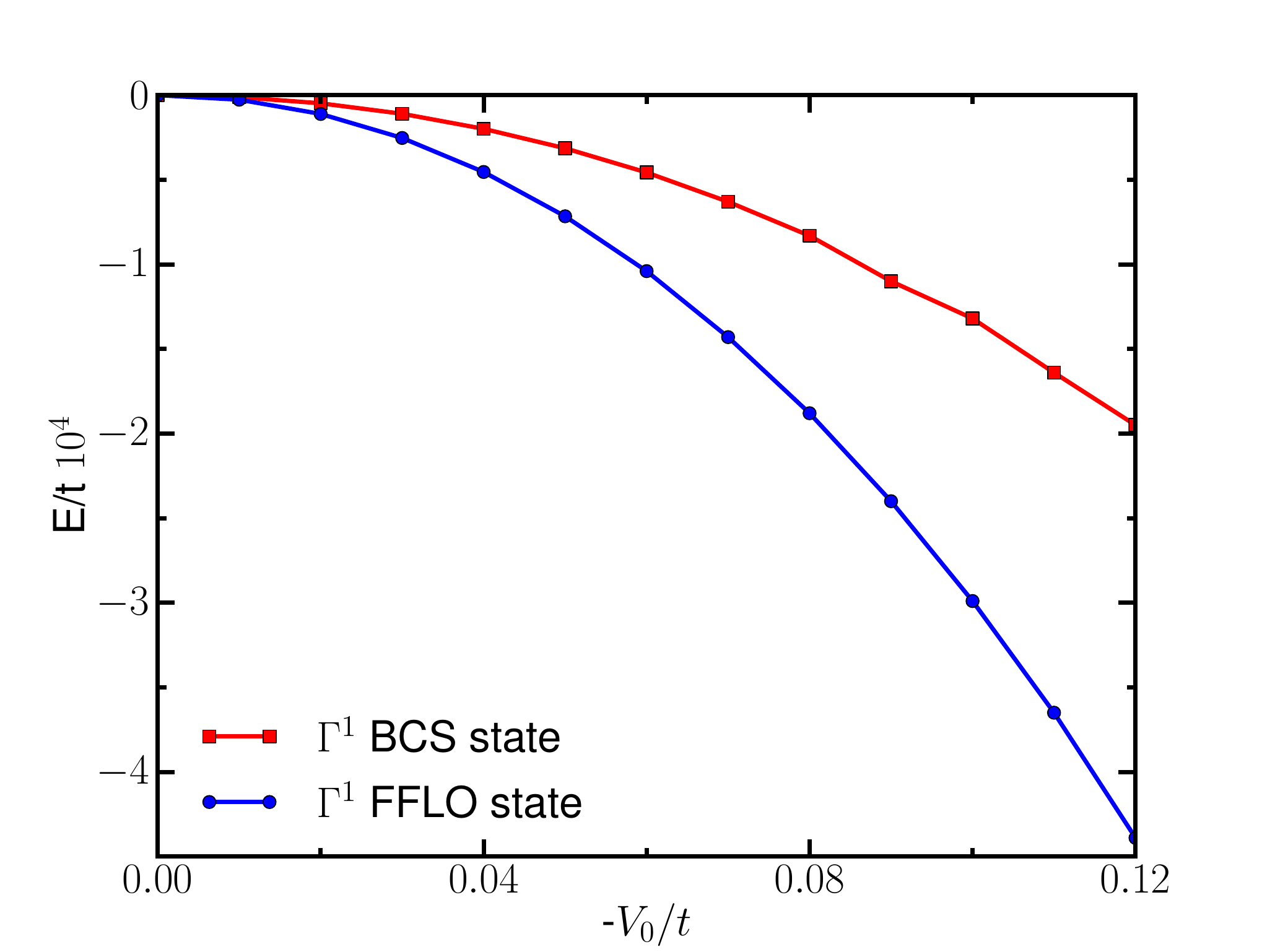}
\caption{Mean-field energy $E$ of the even-parity $\Gamma^{1}$-BCS and the $\Gamma^{1}$- FFLO states as a function of interaction strength $V_0$. Other model parameters used to obtain this plot are $\mu/t = 0.3$, $Q = 0.7$, $V_1 = 0$, and $d\Omega = 0.2$}
\label{Fig2}
\end{figure}

In Fig.~\ref{Fig2} we plot the mean-field energy for the two parings, as obtained from Eq.~\eqref{Energy}, as a function of the interaction strength $V_0$. The $\Gamma^{1}$-FFLO state has a larger gap than the $\Gamma^{1}$-BCS state and is energetically favored. This result can be understood by considering the spin-momentum locking around the Fermi surface. For the even-parity pairing state, the state $|{\vec k}, \alpha\rangle$ ($\alpha$ is the spin state) is paired with the inversion partner state $|-{\vec k}, \sigma^{z}\alpha\rangle$. The pairing amplitude is of the form $\sim \langle c^{\dagger}({\vec k})i\sigma^{y}c^{*}(-{\vec k}) \rangle$ which takes the maximum value if the two states at ${\vec k}$ and $-{\vec k}$ have opposite spins. However, the spins at ${\vec k}$ and $-{\vec k}$ are not anti-parallel and even become parallel at the poles (which is the origin of the nodes, see Fig.~\ref{Fig1}) which tends to reduce the superconducting gap. In contrast, the FFLO state connects the states $|{\vec k}+{\vec Q}, \alpha \rangle$ and $|-{\vec k}+{\vec Q}, \beta \rangle$ via the spin singlet channel with $\beta = -\alpha$ (anti-parallel spins). A gap opens up everywhere at the Fermi surface with a larger gap than the even-parity BCS state. This is very similar to the surface of topological insulator which we discuess Appendix B (See also Ref. [\onlinecite{FFLOsurface}]). The similar finite-momentum pairing (``intra-valley" pairing or ``Kekule" pairing) can happen in a graphene in the presence of a nearest-neighbor attractive interaction~\cite{Gr1, Gr2}.      

\section{Discussion}

In this section, we discuss the nodal structure of the $\Gamma^{1}$-BCS state and the effect of the disorder on the $\Gamma^{1}$-FFLO state. Like in the last section, we consider only the singlet components of these states, assuming $|V_{0}|\gg |V_{1}|$. 

\subsection{$\Gamma^{1}$-BCS state}
\label{BCS state}

The pair potential term in the mean field Hamiltonian of the singlet component of the $\Gamma^1$-BCS state is
\begin{align}
H^{\rm BCS}_{\rm pair} &= \sum_{{\vec k}}\Delta c^{\dagger}_{\alpha}({\vec k}) (i\sigma^{y})^{\alpha\beta} c^{\dagger}_{\beta}(-{\vec k}) + h.c. \\
&= \sum_{\vec q}\Delta \psi^{\dagger}_{a,\alpha}({\vec q})(\tau^{x})^{ab}(i\sigma^{y})^{\alpha\beta} \psi^{\dagger}_{b,\beta}(-{\vec q})+h.c. \notag
\label{singlet1}
\end{align}
The second form is obtained in the low energy theory. This superconducting state is an even-parity state and has four point nodes on the Fermi surface at $q_{x}=q_{y}=0$ and $q_{z} = \pm \sqrt{\Delta^{2}+\mu^{2}}$ (see Fig.~\ref{Fig1}). The gap remains closed even when the triplet pairings of the $\Gamma^{1}$-BCS state in Table~\ref{table1} are included.

The two nodal points near Weyl node ${\vec P}_+$ (${\vec P}_-$) carry a winding number of +1 (-1). To demonstrate this we write down the Bogoliubov-de Gennes (BdG) Hamiltonian $H = \sum_{{\vec k}} \Phi_{{\vec k}} {\tilde H}_{{\vec k}} \Phi_{{\vec k}}$ for $\Phi_{{\vec k}} = (c_{{\vec k}}, i\sigma^{y}c^{*}_{-{\vec k}})^{T}$. In the continuum limit at ${\vec P}_+$ (similar expressions are obtained for ${\vec P}_-$) 
\beq
{\tilde H} =
\begin{pmatrix}
h_{+}({\vec q})& \Delta \sigma^{0} \\
\Delta\sigma^{0}& -h_{-}({\vec q})
\end{pmatrix},
\label{nodes}
\eeq
with $h_{\pm}({\vec q})$ defined in Eq.~\eqref{low}. The quasiparticle spectrum corresponding to this BdG Hamiltonian is 
\beq
E({\vec q}) = \pm [q^{2} + \Delta^{2} + \mu^{2} \pm 2(\Delta^{2}q^{2}_{z} + \mu^{2}q^{2})^{1/2}]^{1/2}
\label{spec}
\eeq
which has nodes at $q_x = q_y = 0$, $q_{z} =\pm \sqrt{\Delta^{2}+\mu^{2}}$, both with chirality of $+1$. Near the nodes $|q_x|,|q_y| \ll |q_z|,|\mu|$, we obtain the anisotropic Weyl spectrum
\beq
E({\vec q}) \approx \pm \left[ (q_{z} \pm \sqrt{\Delta^{2}+\mu^{2}})^{2} + q^{2}_{\perp} (1+ \frac{\mu^{2}}{\mu^{2}+\Delta^{2}})\right]^{1/2},
\eeq
with $q_\perp = (q_x,q_y)$. At zero chemical potential, this is similar to the results of Meng and Balents~\cite{WeylSc} who considered the proximity effect of undoped Weyl semimetals. The effect of nonzero chemical potential is to simply shift the Weyl nodes located at $q_{z} = \pm \Delta$ at $\mu =0$, to $q_{z} = \pm \sqrt{\Delta^{2}+\mu^{2}}$. 

Because of the non-trivial winding number carried by the nodes, the nodal points are robust against small perturbations.  The only way to gap out the nodes is to undergo a pair-annihilation of nodes with the opposite winding numbers, and the nodal points are {\it topologically} stable as long as they are separated enough in momentum space. Strikingly, this nodal structure implies that there will be a zero-energy state on the surface which should be detectable in experiment. This is similar to $^{3}$He-A which is an odd-parity pairing state, while our superconducting phase is realized by the even-parity pairing.

\subsection{$\Gamma^{1}$-FFLO state}
\label{FFLO state}

The singlet component of the $\Gamma^{1}$- FFLO state is fully gapped with a mean-field pair potential 
\beq
H^{\rm FFLO}_{\rm pair} = \Delta c^{\dagger}_{\alpha}({\vec k}+{\vec P}_+)(i\sigma^{y})^{\alpha\beta}c^{\dagger}_{\beta}(-{\vec k}+{\vec P}_+) \pm ({\vec P}_+\leftrightarrow {\vec P}_-)
\label{FFLO}
\eeq
with center-of-momentum of $2{\vec P}_{\pm}$. In the low-energy theory, it can be represented by the {\it intra-node} pairing $\sim \Delta \sum_{{\vec q}}\psi^{\dagger}_{a,\alpha}({\vec q}) (i\sigma^{y})^{\alpha\beta} \psi^{\dagger}_{a,\beta}(-{\vec q})$. 

It is known that some two-dimensional FFLO states with strong spin-orbit coupling and parallel magnetic field are unstable against weak disorder~\cite{disorder1,disorder3}. In contrast, the FFLO state discussed in this paper is found to be robust against weak disorder. In fact, the structure of the FFLO state Eq.~\eqref{fflo} and Eq.~\eqref{FFLO} is more similar to the even/odd-parity state of the doped topological insulators studied in Ref.~\onlinecite{disorder2} than usual FFLO states in the two spatial dimension. This similarity is manifested if we write down the pairing for the Weyl fermions in the continuum limit in the helicity eigenstates
\beq
\Delta_{\pm} \propto e^{i\phi} [\langle \psi_{+}({\vec q})\psi_{+}(-{\vec q}) \rangle \pm \langle \psi_{-}({\vec q})\psi_{-}(-{\vec q}) \rangle],
\label{FFLO2}
\eeq
which corresponds to Eq.~(5) of Ref.~\onlinecite{disorder2}. Within this Cooper channel, we add a scalar disorder term to the Hamiltonian
\beq
H_{\rm imp} = V_{\rm imp} \sum_{{\vec k}, {\vec p} \in FS}c^{\dagger}_{{\vec k},\sigma}c_{{\vec p},\sigma} = \sum_{{\vec q}, {\vec l} \in FS} V^{ab}_{{\vec q}, {\vec l}} \psi^{\dagger}_{a,{\vec q}}\psi_{b,{\vec l}}
\eeq
The matrix element $V^{ab}_{{\vec q}, {\vec l}}$ is given by
\beq
V^{ab}_{{\vec q}, {\vec l}} = V_{imp} 
\begin{pmatrix}
\langle{\hat q}|{\hat l}\rangle& \langle{\hat q} | {\bar l}\rangle  \\
\langle {\bar q}|{\hat l}\rangle & \langle{\bar q}|{\bar l}\rangle
\end{pmatrix},
\label{disorders}
\eeq
where we have used the standard normalized spin state ${\hat q} \cdot {\vec \sigma}|{\hat q}\rangle = |{\hat q}\rangle$ and ${\bar q} \cdot {\vec \sigma} |{\bar q}\rangle = |{\bar q}\rangle$ with ${\hat q} = {\vec q}/|{\vec q}|$ and ${\bar q} = ({\vec q}_{\perp}, -q_{z})/ |{\vec q}|$. With this impurity scattering, the self-energy can be worked out in the self-consistent Born approximation, and we find that the correction to the self-energy and the Cooperon diagram due to disorder are exactly of the same form as obtained by Michaeli and Fu~\cite{disorder2}. In fact, the only difference between the FFLO state $\Delta_{\pm}$ in Eq.~\eqref{FFLO} and the even/odd-parity paired states of Ref.~\onlinecite{disorder2} is phase factors in the matrix elements of $V^{ab}_{{\vec k}, {\vec l}}$ in Eq.~\eqref{disorders} which does not show up in the corrections to the self-energy, the Cooperon diagram, and the pairing susceptibilities. Thus we conclude that the critical temperature of $\Gamma^{1}$- FFLO states is not affected by the disorder and thus $\Gamma^{1}$- FFLO state is robust.

\section{conclusion}

In conclusion, we have studied the possible superconducting states of doped inversion-symmetric Weyl semimetals. We considered a concrete lattice model realizing a Weyl semimetal and found that the FFLO state has a lower energy than the even-parity state if the interaction is phonon-mediated, and the phase is argued to be stable against disorder. Though the even-parity state is less favored in energy than the FFLO state, it interestingly provides an electronic analogue of $^{3}$He-A phase. 

We remark briefly on the implication of our work for superconducting states of Weyl semimetal models other than the one studied in this paper. Among the many proposals for the Weyl semimetal phase, we restrict ourselves to the models based on the topological insulators~\cite{cho, Burkov} with the time-reversal breaking perturbation.
\beq
H = v \tau^{z}{\vec \sigma}\cdot {\vec k}_{\perp} + \tau^{x}k_{z} + m\sigma^{z}
\label{real}
\eeq
The typical symmetry of the model is $I \times C_{n}$ ($\times M$, Mirror symmetry) where $I$ is the inversion symmetry and $C_{n}$ is the $n$-fold lattice rotation symmetry along a certain axis (for the model based on Bi$_{2}$Se$_{3}$~\cite{cho}, we have $n=3$). Due to the strong spin-orbit interaction, the spatial symmetry operation involves the spin/orbital operations, e.g., $I: {\vec k}\rightarrow -{\vec k}$ should involve $\tau^{y}$, and $\tau^{y}H(-{\vec k}) \tau^{y} = H({\vec k})$ (the lattice rotation will involve a spin rotation). Note that these symmetry considerations already manifest the similarity between the realistic model and the simplified model Eq.~\eqref{lattice}, and this similarity becomes much clearer if we go to the low-energy theory of Eq.~\eqref{real}. It is not difficult to confirm that the low-energy theory is identical to Eq.~\eqref{low}, and hence we will have similar superconducting states, FFLO and electronic analogues of $^{3}$He-A, in the more realistic model. Hence, we predict that the superconducting states we found should show up in other proposals for Weyl semimetals.

Note that FFLO state Eq.~\eqref{FFLO} shows a density modulation pinned by the momentum of the Weyl nodes (which is reminiscent of the field-induced charge density wave~\cite{Yang} of the Weyl semimetals). Many experimentally available Weyl semimetals have a large number of Weyl nodes, 
for example the irridates which have 24 nodes~\cite{Ashvin} (or an inversion-symmetry broken Weyl semimetal has {\it at least} four Weyl nodes~\cite{murakamiweyl}). While our minimal model calculation here does not guarantee that the FFLO state will be the lowest energy state in such systems, 
at minimum it suggests that it will be a competing state. In this case the FFLO state can have multiple centers of momenta. This directly implies that there will be interesting density modulation patterns which are fully determined by the position of the Weyl nodes. (This is true at least at the level of mean-field theory which ignores the effect of $O(\Delta^{4})$ terms in the Landau-Ginzburg theory. $O(\Delta^{4})$ terms can potentially {\it melt} this pattern).

We also note that FFLO state can host interesting half-quantum vorticies discussed in Ref. [\onlinecite{VortexFFLO}]. In the FFLO state, we have {\it two} indepedent superconducting order parameters $\Delta(\pm {\vec P}) \propto \exp(\pm 2i{\vec P}\cdot {\vec r})$,i.e., the order parameter space is $S^{1}\times S^{1}$. The half-quantum vortex corresponds to a unit ``winding" of the phase of the $\Delta({\vec P})$ while the phase of the $\Delta(-{\vec P})$ does not wind. On the other hand, the Fermi surface around the Weyl node at $\vec P$ encloses the $\pi$-Berry phase~\cite{Hosur_Vishwanath} which signals that there will be a gapless ``chiral'' Majorana mode at the core of the half-quantum vortex. Furthermore, this implies that a full quantum vortex will be a composite of the two half-quantum vortices and each half-quantum vortex will have a chiral mode. Thus the full quantum vortex will host a helical Majorana mode. In contrast to the related case~\cite{Hosur_Vishwanath}, this helical Majorana mode is not symmetry protected and is therefore generally gapped out. Furthermore, the helical Majorana mode can be understood as the critical point between a weak pairing state and a strong pairing state in a $1$D p-wave superconductor~\cite{Kitaev}. There are two possible phases for the full quantum vortex depending on the sign of ``mass gap'' for the helical mode~\cite{Kitaev} and in a nontrivial phase there will be a Majorana fermion at the end of the vortex.

\acknowledgements
The authors thank Pavan Hosur, Ashvin Vishwanath, Sid Parameswaran, Eun Gook Moon, Yong Baek Kim, and Tarun Grover for helpful discussion and Daniel Agterberg for useful comments from which we learn about the half-quantum vortex in FFLO states. The authors acknowledge support from NSF DMR-1206515 (G. Y. C. and J. E. M.), Office of BES, Materials Sciences Division of the U.S. DOE under contract No.DE-AC02-05CH1123 (Y. M. L.), and the LBNL Thermoelectrics Program (J. H. B.) of DOE BES.

\appendix
\section{Continuum Weyl fermions: interaction, and gap equation}
\label{app:interactions}
In this appendix, we derive the equations related to the continuum theory from the microscopic lattice model Eq.~\eqref{lattice}. First of all, let us derive the interaction Eq.~\eqref{ContinuumInteraction} for the Weyl fermions. The interaction \eqref{interaction} in the momentum space is
\beq
H = \sum_{{\vec k},{\vec p}, {\vec q}}V_{{\vec k}} c^{\dagger}_{\sigma}({\vec k}+{\vec p})c^{\dagger}_{\tau}({\vec q}-{\vec k})c_{\tau}({\vec q})c_{\sigma}({\vec p}),
\eeq
with $V_{{\vec k}} = V_{0} + V_{1} (\cos(k_{x}) + \cos(k_{y}) + \cos(k_{z}))$. For the BCS pairings where the center of momentum is at zero ${\vec q} = - {\vec p}$, we obtain the pairing potential
\beq
H = \sum_{{\vec p}, {\vec q}}V_{{\vec p}-{\vec q}} c^{\dagger}_{\sigma}({\vec p})c^{\dagger}_{\tau}(-{\vec p})c_{\tau}(-{\vec q})c_{\sigma}({\vec q}).
\label{int}
\eeq
As the electron operators that we are concerning are localized near the Weyl points, we expand the electron operators near the Weyl points. This can be easily done by plugging ${\vec p} = {\vec k}\pm{\vec P}$ and ${\vec q} = {\vec l} \pm {\vec P}$ with ${\vec P} = (0,0,Q)$ into Eq.~\eqref{int}. For example, the interaction \eqref{int} includes the interaction
\begin{align}
\sim&  V_{({\vec k}+{\vec P})- ({\vec l}+{\vec P})} \psi^{\dagger}_{+,\sigma}({\vec k})\psi^{\dagger}_{-,\tau}(-{\vec k})\psi_{-,\tau}(-{\vec l})\psi_{+,\sigma}({\vec l})\nonumber\\
=& V^{+-,-+}({\vec k}-{\vec l})\psi^{\dagger}_{+,\sigma}({\vec k})\psi^{\dagger}_{-,\tau}(-{\vec k})\psi_{-,\tau}(-{\vec l})\psi_{+,\sigma}({\vec l}),
\end{align}
which allows us to identify $V^{+-,-+}({\vec k}-{\vec l}) = V_{({\vec k}+{\vec P})- ({\vec l}+{\vec P})}$. Similarly, we can identify $V^{-+,+-}({\vec k}-{\vec l}) = V_{({\vec k}-{\vec P})- ({\vec l}-{\vec P})}$, $V^{-+,-+}({\vec k}-{\vec l}) = V_{({\vec k}-{\vec P})- ({\vec l}+{\vec P})}$, and $V^{+-,+-}({\vec k}-{\vec l}) = V_{({\vec k}+{\vec P})- ({\vec l}-{\vec P})}$. After this identification, it is straightforward to expand for the small ${\vec k}, {\vec l}$ to obtain
\begin{align}
&V^{-+;+-}=V^{+-;-+}= V_{0} + 3V_{1} - \frac{V_{1}}{2}({\vec k}-{\vec l})^{2}, \nonumber\\
&V^{-+;-+}= V_{0}+2V_{1} +V_{\perp} + V^{+}_{\parallel},\nonumber\\
&V^{+-;+-} =V_{0}+2V_{1} +V_{\perp} + V^{-}_{\parallel}, \nonumber \\
&V_{\perp} = -\frac{V_{1}}{2} ({\vec k}_{\perp}-{\vec l}_{\perp})^{2},\nonumber\\
&V^{+}_{\parallel} = V_{1} (\cos(2Q) (1 - \frac{1}{2}(k_{z}-l_{z})^{2}) + (k_{z}-l_{z})\sin(2Q) ),\nonumber \\
&V^{-}_{\parallel} =V_{1} (\cos(2Q) (1 - \frac{1}{2}(k_{z}-l_{z})^{2}) - (k_{z}-l_{z})\sin(2Q) ).
\end{align}
Next, we discuss the mean-field pairing terms in the continuum theory derived from the lattice model. The possible pairing states from the interaction Eq.~\eqref{interaction} are listed in the table \ref{table1}. The typical form of the pairing term can be represented by
\beqn
&\notag H_{\rm pair} = \sum_{{\vec k}}\Delta c^{\dagger}({\vec k}) \Gamma({\vec k})c^{\dagger}(-{\vec k})\\
&=  \sum_{{\vec k}}\Delta c^{\dagger}_{\sigma}({\vec k}) \Gamma_{\sigma\tau}({\vec k})c_{\tau}^{\dagger}(-{\vec k}).
\eeqn
(that is, $\Delta_{\sigma\tau}({\vec k}) = \Delta \Gamma_{\sigma\tau}({\vec k})$). As we did for the interaction terms, we expand the pairing terms near the Weyl points by
\begin{align}
& c^{\dagger}_{\sigma}({\vec k}) \Gamma_{\sigma\tau}({\vec k})c_{\tau}^{\dagger}(-{\vec k})\nonumber\\
&= \psi^{\dagger}_{+,\sigma}({\vec p}) \Gamma^{+-}_{\sigma\tau}(+,{\vec p})\psi^{\dagger}_{-,\tau}(-{\vec p})\nonumber\\
\Gamma_{2} &= (p_x+ip_y)\tau^x(\sigma^0+\sigma^z),(p_x-ip_y)\tau^x(\sigma^0-\sigma^z); \nonumber\\
\Gamma_{3, +} & = (p_x+ip_y)\tau^y\sigma^y,i\tau^y(\sigma^0+\sigma^z); \nonumber\\
\Gamma_{3,-} &= (p_x-ip_y)\tau^y\sigma^y,i\tau^y(\sigma^0-\sigma^z).
\label{Pair}
\end{align}

As we now have the explicit form of the pairing term and the interaction for the Weyl fermions, it is now straightforward to obtain the gap equation for each ansatz by solving
\beq
\Delta^{ab}_{\sigma\tau}({\vec p}) = \sum_{{\vec k}}V^{ab;cd}({\vec p} - {\vec k}) \langle \psi_{c,\tau}(-{\vec k})\psi_{d,\sigma}({\vec k}) \rangle.
\label{eqnn}
\eeq
For example, the gap equation for the singlet pairing component $\Delta^{ab}_{\alpha\beta} = \Delta (\Gamma^{1})^{ab}_{\alpha\beta}$ in Eq.~\eqref{Pair} is
\beqn
&\notag\Delta = \frac{1}{4} \sum_{{\vec k}}(\tau^{x})_{ab}V^{ab;cd}({\vec p} - {\vec k})\cdot\\
& \langle \psi_{c,\tau}(-{\vec k})(-i\sigma^{y})^{\tau\sigma}\psi_{d,\sigma}({\vec k}) \rangle.
\eeqn
The similar expressions of the gap equations hold for other pairings and the results are following. For the triplet pairing component in $\Gamma^1$ ($C_{4}=1$), there are two pairing channels $\sim \Delta_{I}(X-iY)(\sigma^{x}+i\sigma^{y})(i\sigma^{y}) +  \Delta_{II}(X+iY)(\sigma^{x}-i\sigma^{y})(i\sigma^{y})$ which can mix each other (see table \ref{table1})
\begin{align}
\Delta_{I} &= \frac{1}{8} \sum_{{\vec k}}\{[V^{x;cd}({\vec k})-iV^{y;cd}({\vec k})][X_{cd}({\vec k})+iY_{cd}({\vec k})]\}, \nonumber \\
\Delta_{II} &= \frac{1}{8} \sum_{{\vec k}}\{[V^{x;cd}({\vec k})+iV^{y;cd}({\vec k})][X_{cd}({\vec k})-iY_{cd}({\vec k})]\},
\end{align}
where we define the following compact notations
\begin{align}
V^{x,cd} ({\vec k})&= \frac{\sum_{\vec p \in d\Omega} (\tau^{x})_{ab} p_{x}V^{ab,cd}({\vec k}-{\vec p})}{\sum_{\vec p \in d\Omega} p^{2}_{x}} \nonumber \\
V^{y,cd} ({\vec k})&= \frac{\sum_{\vec p \in d\Omega} (\tau^{x})_{ab} p_{y}V^{ab,cd}({\vec k}-{\vec p})}{\sum_{\vec p \in d\Omega} p^{2}_{y}}
\end{align}
where $d\Omega$ is the thin shell around the Fermi surface. The width of the shell is determined by the phenomenological parameter (Debye frequency) defining the electron-phonon coupling, e.g.,$t \times d\Omega$ is the characteristic energy of the phonons.
\begin{align}
X_{cd}({\vec k}) &= \langle \psi_{c,\beta}(-{\vec k}) [-i\sigma^{y}\sigma^{x}]^{\beta\alpha}\psi_{d,\alpha}({\vec k})\rangle \nonumber\\
Y_{cd}({\vec k}) &= \langle \psi_{c,\beta}(-{\vec k}) [-i\sigma^{y}\sigma^{y}]^{\beta\alpha}\psi_{d,\alpha}({\vec k}) \rangle \nonumber \\
Z_{cd}({\vec k}) &= \langle \psi_{c,\beta}(-{\vec k}) [-i\sigma^{y}\sigma^{z}]^{\beta\alpha}\psi_{d,\alpha}({\vec k})\rangle
\end{align}
with the expectation value taken for the mean-field superconducting state.

For $\Gamma^{2}$ pairing ($C_{4}=-1$), there are two pairing channels $\sim \Delta_{I}(X+iY)(\sigma^{x}+i\sigma^{y})(i\sigma^{y}) +  \Delta_{II}(X-iY)(\sigma^{x}-i\sigma^{y})(i\sigma^{y})$ which can mix each other (see table \ref{table1}), and the self-consistency requires
\begin{align}
\Delta_{I} &= \frac{1}{8} \sum_{{\vec k}}\{[V^{x;cd}({\vec k})+iV^{y;cd}({\vec k})][X_{cd}({\vec k})+iY_{cd}({\vec k})]\}, \nonumber \\
\Delta_{II} &= \frac{1}{8} \sum_{{\vec k}}\{[V^{x;cd}({\vec k})-iV^{y;cd}({\vec k})][X_{cd}({\vec k})-iY_{cd}({\vec k})]\},
\end{align}
For $\Gamma^{3,\pm}$ pairing ($I= -1$ and $C_{4} =\pm i$), the analoguous expressions can be derived by similar methods.

\section{FFLO state on Surface of topological insulators}
In this appendix, we will propose a possible FFLO state from the surface state of the strong topological insulator under the parallel magnetic field. As far as the magnetic field is in-plane, the only coupling from the magnetic field to the surface state is Zeeman coupling. Then, the Zeeman coupling will shift the Fermi surface uniformly along the direction to the magnetic field. The low-energy theory is
\beq
H = v{\vec \sigma}\cdot ({\vec k} + g{\vec B}/v) - \mu.
\eeq
If there is a phonon-mediated {\it surface} transition toward superconducting state, the superconducting state should be at the finite center-of-momentum pairing. We model the phonon-mediated attractive interaction as
\beq
\delta H = U \sum_{{\vec k}} n_{\vec k}n_{-{\vec k}},
\eeq
which is local in the real space and uniform in the momentum space. This implies that the pairing will be uniform in the momentum space, hence we can single out a single pairing state $\Delta({\vec r}) \propto \exp(-2i{\vec Q}\cdot {\vec r})$ with ${\vec Q} = g{\vec B}/v$,
\beq
H_{\rm pair} = \Delta({\vec Q}) c^{\dagger}_{\alpha}({\vec Q}+{\vec k}) (i\sigma^{y})^{\alpha\beta}c^{\dagger}_{\beta}({\vec Q}-{\vec k}) + h.c.
\eeq
Hence, we have shown that FFLO state can show up, at least in the mean-field theory, in the surface state of strong topological insulators under the in-plane magnetic field. What would be the effect of weak disorder to this phase? By following the discussion in reference [\onlinecite{disorder1}], we conclude that this FFLO state should be robust against the weak neutral disorder (this problem corresponds to the problem where the disorder scatters electrons only within the {\it single} Rashba band in the reference).

\section{Symmetry classification of pairing order parameters in two-band model (\ref{lattice})}
\label{app:symm}
\begin{table}[tb!]
\begin{tabular} {c|c|c|c|c}
\hline
$IRR$&$C_{4}$&basis functions of $\Delta({\vec{k}})$&Nodes?\\ \hline
$A_{u}$&$1$&$Z$&N/A\\ \hline
$B_{u}$&$-1$&$XYZ~\text{or}~(X^2-Y^2)Z$&line nodes\\ \hline
$E_{u+}$&$i$&$X+iY$&N/A\\ \hline
$E_{u-}$&$-i$&$X-iY$&N/A\\
\hline
\end{tabular}
\caption{Symmetry classification of distinct BCS pairing order parameters in (\ref{BCS pair}), corresponding to different odd-parity irreducible representations (IRRs) of point group $C_{4h}$. Here $(X, Y, Z)$ are basis functions for the momentum-space pairing function $\Delta(\vec{k})$, denoting e.g. the function $(\sin(p_{x}),\sin(p_{y}),\sin(p_{z}))$ or other functions with the same symmetry.}
\label{tab:C4h BCS}
\end{table}
In this section we classify possible BCS-type pairing order parameters in two-band model (\ref{lattice}) with interactions, as long as interaction terms do not break the point group symmetry $C_{4h}$ of tight-binding model (\ref{lattice}). These different pairing order parameters are characterized by distinct irreducible representations of group $C_4$ generated by 90 degree rotation along $\hat{z}$-axis. 

When the Weyl semimetal described by two-band model (\ref{lattice}) is doped slightly with $\mu>0$ and $|\mu/t|<<1$, the Fermi surface consists two electron pockets around Weyl nodes $\pm\vec{P}=(0,0,\pm Q)$. The tight-binding model (\ref{lattice}) can be diagonalized into $H(\vec{k})=\vec{d}_{\vec{k}}\cdot\vec{\sigma}-\mu=U_{\vec{k}}^\dagger(\varepsilon_{\vec{k}}\sigma^z-\mu)U_{\vec{k}}$, where $U_{\vec{k}}=\begin{pmatrix}u_{\vec{k}}&v_{\vec{k}}\\-v_{\vec{k}}^\ast&u^\ast_{\vec{k}}\end{pmatrix}$ is a $2\times2$ unitary matrix. Hence the low-energy degree of freedoms on the Fermi surface $\varepsilon_{\vec{k}}=\mu$ are
\beq
f_{\vec{k}}=u_{\vec{k}}c_{\vec{k},\uparrow}+v_{\vec{k}}c_{\vec{k},\downarrow},
\label{eigenband}
\eeq
and according to inversion symmetry we have $U_{-\vec{k}}=U_{\vec{k}}\sigma_{z}$ and 
\beq
f_{-\vec{k}}=u_{\vec{k}}c_{-\vec{k},\uparrow}-v_{\vec{k}}c_{-\vec{k},\downarrow}.
\eeq

For a general electronic model with $C_{4h}$ symmetry (but no time reversal), the Fermi surface is nondegenerate. Therefore a generic BCS-type pairing term is written as
\beq\label{BCS pair}
H_{\rm pair}=\sum_{\vec{k}}\Big(\Delta({\vec{k}})f^\dagger_{\vec{k}}f^\dagger_{-\vec{k}}+~h.c.\Big),
\eeq
Apparently the pairing order parameter $\Delta(-\vec{k})=-\Delta({\vec{k}})$ is a odd-parity function (instead of a $2\times2$ matrix). One can always choose a gauge so that under $C_4$ symmetry in (\ref{symmetry}):
\beq
C_4:~~~(k_x,k_y,k_z)\rightarrow(k_y,-k_x,k_z)
\eeq
the eigenvector $(u_{\vec{k}},v_{\vec{k}})$ transforms as 
\beq
u_{C_4\vec{k}}=u_{\vec{k}},~~~v_{C_4\vec{k}}=i\cdot v_{\vec{k}}.
\eeq
In this specific gauge the operator $f_{\vec{k}}$ transforms trivially under rotation $C_4$ and inversion $I$, hence the order parameter $\Delta(\vec{k})$ should form a one-dimensional odd-parity (irreducible) representation of the symmetry group $C_{4h}$. All possible different order parameters are classified by their symmetry and listed in Table \ref{tab:C4h BCS}. The $B_u$ state has nodal lines along which there are gapless excitations, while the other three BCS paired states are fully gapped.

The symmetry classification of BCS pairing order parameters are generally true for any system with $C_{4h}$ symmetry but no time reversal (so that the Fermi surface is nondegenerate). In the specific two-band model (\ref{lattice}), we can relate the low-energy eigenband operators $f_{\vec{k}}$ to original electron operators $c_{\vec{k},\sigma}$ through (\ref{eigenband}), and the distinct pairing functions in the original electron basis are summarized in Table \ref{table1}.

\bibliography{WeylSC_updated}
\end{document}